\documentclass[dvipsnames,11pt]{article}
\usepackage{adjustbox}
\usepackage{amssymb}
\usepackage{amsthm}
\usepackage{amsmath}
\usepackage{authblk}
\usepackage{blkarray}
\usepackage{booktabs}
\usepackage{bm}
\usepackage{calc}
\usepackage{caption}
\usepackage{cases}
\usepackage{centernot}
\usepackage[utf8]{inputenc}
\usepackage{dcolumn}
\usepackage{float}
\usepackage[margin = 1in]{geometry}
\usepackage{graphicx}
\usepackage{lineno}
\usepackage{listings}
\usepackage{makecell}
\usepackage{mathtools}
\usepackage[numbers]{natbib}
\usepackage{pifont}

\usepackage{setspace}
\usepackage{soul}
\usepackage{subcaption}
\usepackage{tabularx}
\usepackage{tgpagella}
\usepackage{verbatim}
\usepackage{xcolor}
\usepackage{xfrac}
\usepackage[style=english]{csquotes}
\MakeOuterQuote{"}
\title{Social and individual learning in the Minority Game}
\author[1,2,*]{Bryce Morsky}
\author[1]{Fuwei Zhuang}
\author[1]{Zuojun Zhou}
\affil[1]{Department of Mathematics \& Statistics, Queen's University, Kingston, ON, CA}
\affil[2]{Department of Mathematics, Florida State University, Tallahassee, FL, USA}
\affil[*]{Corresponding author: bmorsky@fsu.edu}
\date{\today}

\begin{document}

\maketitle

\begin{abstract}
    We study the roles of social and individual learning on outcomes of the Minority Game model of a financial market. Social learning occurs via agents adopting the strategies of their neighbours within a social network, while individual learning results in agents changing their strategies without input from other agents. In particular, we show how social learning can undermine efficiency of the market due to negative frequency dependent selection and loss of strategy diversity. The latter of which can lock the population into a maximally inefficient state. We show how individual learning can rescue a population engaged in social learning from such inefficiencies.
\end{abstract}


\section{Introduction}

The Minority Game, inspired by the El Farol Bar problem \cite{arthur94}, is a well-studied model of financial markets that exhibits self-organization of agents' decisions and phase transitions \cite{challet08,challet01a,challet01b,challet97,savit99,yeung12}. It models a simplified market of buyers and sellers who use the history of past outcomes of the game to predict future ones. An agent's action --- whether to buy or sell --- will be successful if their choice is the minority action of the population: the forces of supply and demand on prices being the intuition behind this. A popular approach to modelling the Minority Game is to let agents have \emph{strategy tables} that predict the future minority action based on the history of minority actions and thus provide recommendations. Using these strategy tables and given the history of minority actions, agents choose whether to buy or sell each round of the game. Since the size of the minority does not matter, there is an issue of efficiency in the market. The larger the minority, the more agents win in that round. Though a simplified representation of a market, this model is able to exhibit many qualitative behaviours of real markets \cite{challet01b,ferreira03,ortisi13}.

There have been many extensions to the strategy table model of the Minority Game, in particular the Evolutionary Minority Game \cite{johnson99,li00a,li00b,lo00,yang03}. In this framework, agents have an evolvable trait that determines whether or not they deviate from the strategies that the predictions of their strategy tables would recommend. The model has been extended in several ways, such as by incorporating a "genetic crossover" algorithm in which poorly performing agents can generate new strategies from their current strategies \cite{sysi03}, and by setting the agents on networks \cite{burgos04,fagiolo05,shang07,zhang18}. Evolution in the Minority Game on networks occurs by agents imitating neighbouring agents by copying their behaviours \cite{chen09,hod04,lavicka07,shang06,shang07,slanina00,slanina01}. A variety of networks have been explored such as complete, random, regular, and scale-free networks. Imitation on these networks can be based on the payoffs of neighbours, and can lead to herding behaviour and thus high volatility \cite{cajueiro06}. Theoretical studies of the impact of social networks of peers have also shown that they can delay stabilization of markets \citep{bakker10}.

Our goal is to extend this line of research by further infusing ideas from social and individual learning \cite{feldman96,vriend00} into the Minority Game to explore the effects of learning others' and developing novel investment strategies. Social learning is a mechanism of cultural transmission that has been well studied in biology and anthropology \cite{laland01}. Through it, cultural traits, in our case financial strategies, can evolve in a population. Our framework thus fits under the novel field of social finance \cite{kuchler21}, which crosses the fields of finance with that of cultural evolution, the evolutionary study of social change \cite{boyd88, cavalli81, henrich03}. It aims to understand markets through an ecological, evolutionary, and psychological view \cite{akcay21,farmer99, hirshleifer20}. Social finance considers the ecology of investors: how they adopt beliefs and investment strategies from one another and how those impact market prices. Under the adaptive markets hypothesis \cite{lo00}, the change in investor behaviour is analogous to ecological and evolutionary processes in biology: through selection and adaptation, investors change their strategies in an evolving market.

Here, we explore the effects of the evolution of financial strategies in a framework of social finance. We do not consider the same evolving trait as in the Evolutionary Minority Game \cite{johnson99}. Rather, we consider the case where individuals can learn new strategy tables either via imitating their neighbours on a social network who have higher payoffs (i.e.\ social learning), or via creating novel strategy tables (i.e.\ individual learning). Such payoff-based social learning is well established in the literature on cultural evolution \cite{kendal09}. With respect to financial markets, investing advice can spread on social networks of investors via social networks \citep{koochakzadeh12,simon12,tu18,li21}. Examples include social media platforms such as SeekingAlpha, StockTwist \citep{tu18}, and Facebook \citep{simon12}. Although, a key difficulty for investors is extracting valuable information from such networks \citep{wang14}.

Connections between mutual fund managers and corporate board members, often developed during individuals' education, can also influence investment decision making and lead to similar investment strategies \cite{cohen08,fracassi17}. Since such social influences of investment decision making are prevalent and significant, we believe that it is important to incorporate social learning into models such as the Minority Game.

We contrast social learning with individual learning, which models how investors can create novel investment strategies independent of their social networks. Whether these are developed via trial and error or some other means, they are not correlated with the strategies of those in their social network. In comparing models with varying degrees of social and individual learning and different social networks, we explore how these factors impact the efficiency of the market as modeled by the Minority Game. We show how social learning can undermine the efficiency of markets, and how individual learning can mitigate this effect.

\section{Methods}

Consider a population of $N$ agents playing the Minority Game. Each agent has $S$ strategy tables numbered $1$ through $S$, which are initially assigned randomly. These strategy tables are dictionaries with different combinations of memories of the last $M$ winning actions as keys and recommended actions as values: "1" or "-1" or, equivalently, "buy" or "sell". Memory length $M$ is sometimes considered the "brain size" of agents, and each strategy table will thus have $2^M$ recommendations. Table \ref{table:strategy_table} represents a strategy table with the key as a string of 1's and -1's and for $M=3$. Agents must choose one of their strategy tables and follow its recommendation independently of others' actions each turn.

\begin{table}[!htpb]
\begin{center}
\begin{tabular}{cc}
\toprule
Memory (key) & Recommendation \\
\midrule
111 & 1 \\
11-1 & -1 \\
1-11 & 1 \\
1-1-1 & 1 \\
-111 & 1 \\
-11-1 & -1 \\
-1-11 & -1 \\
-1-1-1 & 1 \\
\bottomrule
\end{tabular}
\end{center}
\vspace{-2mm}
\caption{A strategy table for $M=3$ where "1" and "-1" are buy and sell, respectively. Memories of the past three turns are represented by a string of the minority strategy on turns $n-3$, $n-2$, and $n-1$ for current turn $n$ in that order.} \label{table:strategy_table}
\end{table}

The action chosen by agent $i$ at time $t$ is denoted by $a_{i}(t)$. At the end of each round, we count the number of buys and sells and whichever side has more than $50\%$ of the total will be considered a losing strategy and the other strategy will be the winner. The attendance $A(t)$ is the total number of buys agents make minus the number of sells in a given turn $t$: it's thus a measure of the collective actions of all agents. Explicitly, it is:
\begin{equation}
    A(t) = \sum_{i=1}^N a_{i}(t).
\end{equation}
A positive attendance indicates that the sell or $-1$ action prevails as the winning choice for that round. Conversely, a negative attendance value signifies that the buy or $1$ prevails.

During a turn of the game, all strategy tables earn virtual payoffs, which are payoffs an agent would have earned had they employed their recommendation. We let $\pi_{ik}(t)$ be the virtual payoff for player $i$'s $k$th strategy table at time $t$. Agents will follow the recommendation of their strategy table with the greatest virtual payoff. In the event of a tie, the lower numbered strategy table is followed. Real payoffs are earned by agents who choose a specific recommendation from a strategy table and engage in it. Real payoffs of agents with the losing strategy will be reduced by one, while the real payoffs of agents with the winning strategy will be increased by one. Virtual payoffs are calculated similarly: each strategy table earning them as if their recommendation that turn had been chosen. Specifically, the virtual payoff of the $k$th strategy for agent $i$ evolves by:
\begin{equation}
    \pi_{ik}(t + 1) = \pi_{ik}(t) - \text{sign}(a_{i}(t) A(t)),
\end{equation}
where $\text{sign}(\cdot)$ is the sign function. The game continues in this fashion for a number of turns, and agents may switch between different strategy tables as their virtual payoffs vary.

Here we explore an agent-based extension of this game with social and individual learning, akin to selection based on payoff differences and mutation in biology. After every turn, agents will learn socially or individually with probabilities $\ell^s$ and $\ell^i$, respectively. With probability $\ell^s$, an agent will be paired with a different randomly chosen agent to learn from. We consider the case where this imitation occurs across a social network. Specifically, we construct complete graphs and Erd\H{o}s-R\'{e}nyi random graphs where agents can only socially learn from agents with which they share an edge. To construct the Erd\H{o}s-R\'{e}nyi random graphs, a pair of agents is connected with probability $p=d/N$, where $d$ is the mean node degree. Once the graph is constructed, we assume that it remains fixed throughout the game. Once a focal agent has been paired with one of its neighbours on the social network to learn from, we simulate social learning using a logistic function (as used in previous research on social learning and evolutionary game theory literature where it is sometimes called Fermi selection \cite{szabo98,traulsen07,mcelreath08,morsky19}):
\begin{equation}
    P_{i\to j}(t) = \frac{1}{1 + \exp(\kappa(\min(\pi_{ik}(t))-\max(\pi_{jk}(t))))},
\end{equation}
where $P_{i\to j}(t)$ is the probability that the focal agent $i$ imitates a strategy table from its paired agent $j$ at time $t$. More specifically, agent $i$ compares its worst strategy, the one with the lowest virtual payoff $\min(\pi_{ik}(t))$ at that time, with agent $j$'s best strategy, the one with the highest virtual payoff $\max(\pi_{jk}(t))$ at that same time. $\kappa>0$ is the sensitivity to the difference in payoffs between the agents. If imitation occurs, agent $i$ replaces its worst strategy table with agent $j$' best strategy table along with copying the virtual payoff for that strategy. For individual learning, with probability $\ell^i$, we randomly select an agent's strategy table, irrespective of its virtual payoff, and replace it with a randomly chosen strategy table from the set of all possible strategy tables with the same memory length (note this set contains $2^{2^M}$ strategy tables). The new strategy table is initialized with a virtual payoff of zero. 

\begin{table}[!htp]
\begin{center}
\begin{tabular}{ll}
\toprule
Parameter & Definition \\
\midrule
$\alpha = 2^M/N$ & control parameter \\
$d$ & mean node degree \\
$\kappa = 100$ & payoff differential sensitivity \\
$\ell^i$ & individual learning rate \\
$\ell^s$ & social learning rate \\
$M$ & memory length \\
$N$ & number of agents \\
$p=d/N$ & probability of connecting two agents \\
$S$ & strategy tables per agent \\
\bottomrule
\end{tabular}
\end{center}
\vspace{-2mm}
\caption{Summary definitions of parameters with default values.} \label{param}
\end{table}

$\alpha=2^M/N$ is an important control parameter in Minority Games, since the macroscopic behaviour of the system depends on it \cite{savit99}. Thus, we plot much of the outcomes of our simulations with respect to $\alpha$. In our simulations, we track key measures of the outcome of the game: the attendance, volatility, and entropy. The volatility $\sigma^2/N$ is the time average variance of the attendance after each round of the game $\sigma^2$ normalized by the population size $N$.
It is an inverse measure of the efficiency of resource distribution in the game and thus our key measure of the efficiency of the market. The lower it is, the more efficient the system. 
In addition to attendance and volatility, we evaluate the diversity of strategy tables in the population by calculating the Shannon entropy of the strategy tables:
\begin{equation}
    H(s) = -\sum_i s_i \ln(s_i),
\end{equation}
where $s_i$ is the frequency of strategy table $i$ in the population. We track the average of these values over time and over multiple games. A summary of the parameters their default values and the variables tracked can be found in Table \ref{param}.

\section{Results}

\begin{figure}[!ht]
\captionsetup[subfigure]{justification=centering}
    \centering
    \begin{subfigure}[]{0.49\columnwidth}
        \caption{No social or individual learning, $N=1001$}
        \includegraphics[width=\columnwidth]{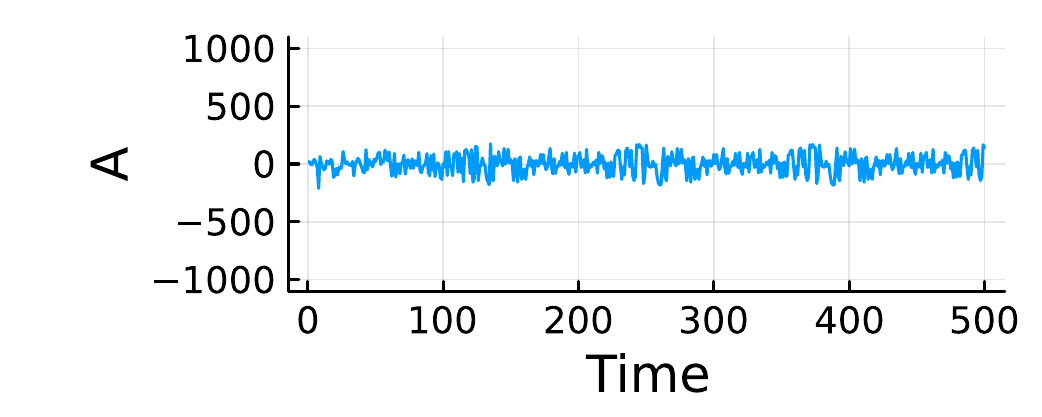}
        \label{fig:ts_M6_N1001_S2}
    \end{subfigure}
    \begin{subfigure}[]{0.49\columnwidth}
        \caption{Only individual learning, $N=1001$}
        \includegraphics[width=\columnwidth]{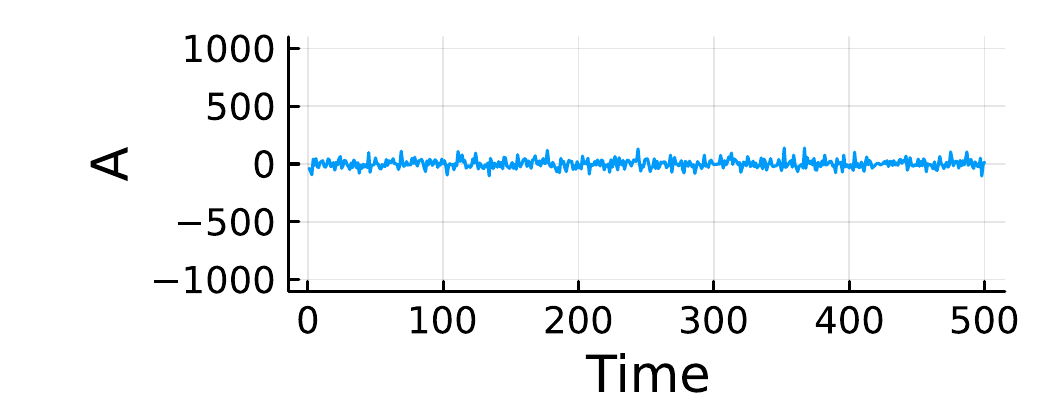}
        \label{fig:ts_ind_learn_li10_M6_N1001_S2}
    \end{subfigure} \\
    \begin{subfigure}[]{0.49\columnwidth}
        \caption{Only social learning, $N=1001$}
        \includegraphics[width=\columnwidth]{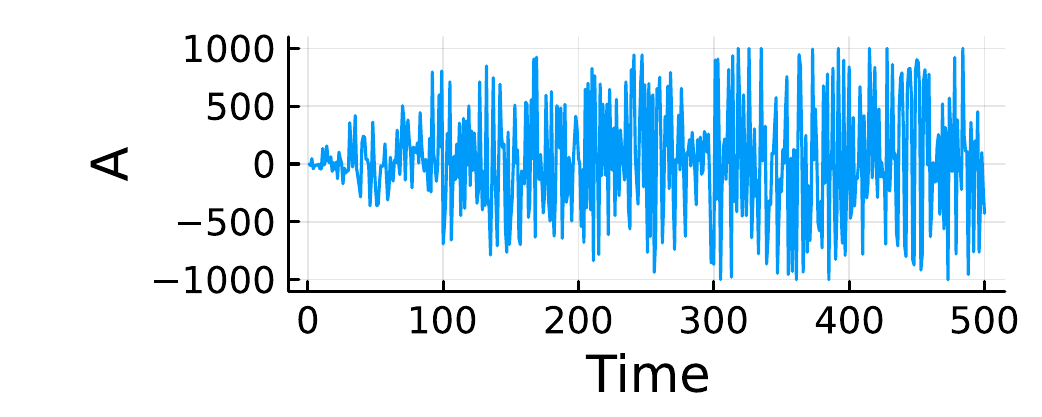}
        \label{fig:ts_soc_learn_ls10_M6_N1001_S2}
    \end{subfigure}
    \begin{subfigure}[]{0.49\columnwidth}
        \caption{Individual and social learning, $N=1001$}
        \includegraphics[width=\columnwidth]{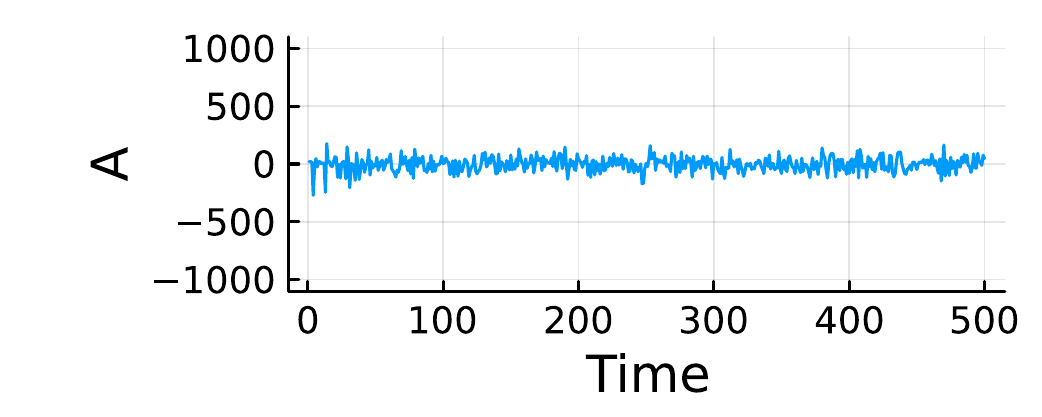}
        \label{fig:ts_ind_soc_learn_lils10_M6_N1001_S2}
    \end{subfigure} \\
    \begin{subfigure}[]{0.49\columnwidth}
        \caption{Only social learning, $N=101$}
        \includegraphics[width=\textwidth]{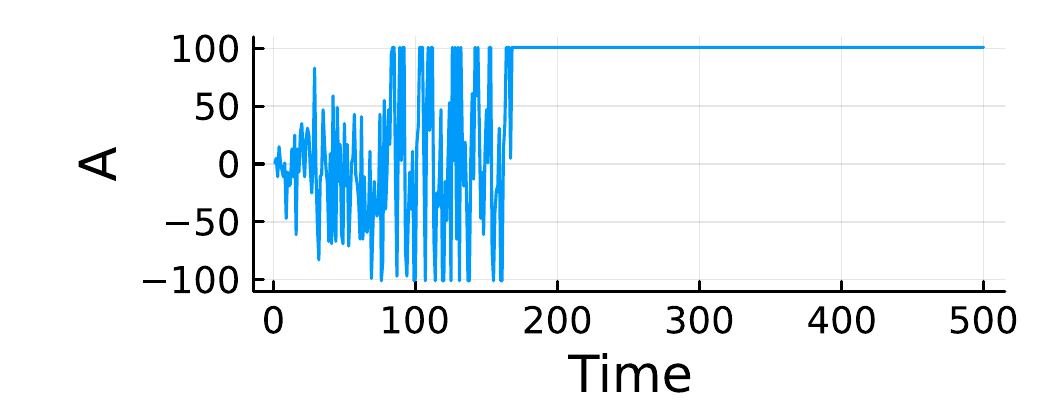} \label{fig:ts_soc_learn_ls20_M6_N101_S2_crash}
    \end{subfigure}
    \caption{Time series of the attendance $A$ for no individual or social learning (a), only individual learning (b), only social learning (c,e), and both individual and social learning (d) on a complete graph (i.e.\ agents can imitate any other agent in the population). $M=6$ and $S=2$. For panel e, $\ell^s=0.2$ and $N=101$, otherwise $\ell^i=\ell^s=0.1$, $N=1001$, and $S=2$.}
    \label{fig:timeseries}
\end{figure}

We begin our results by depicting several informative time series of the attendance $A(t)$ in Figure \ref{fig:timeseries}. We find that when there is only individual learning, the variability in attendance is marginally lower than in the case with no social or individual learning (Figure \ref{fig:ts_ind_learn_li10_M6_N1001_S2}). Social learning, however, on its own results in large swings in the population from buying to selling and thus a high volatility relative to the other cases (Figure \ref{fig:ts_soc_learn_ls10_M6_N1001_S2}). Because, payoffs are negative frequency dependent. A good strategy is readily imitated, which undermines its effectiveness. Individual learning, however, can resolve this issue. Figure \ref{fig:ts_ind_soc_learn_lils10_M6_N1001_S2} depicts a representative example showing that the addition of individual learning reduces the volatility to levels comparable to the case where there is no learning (Figure \ref{fig:ts_M6_N1001_S2}). Without individual learning, the high volatility of social learning can result in the population becoming locked into either buying or selling in which case no agent can ever win. Figure \ref{fig:ts_soc_learn_ls20_M6_N101_S2_crash} is representative of this scenario. The population gets locked into always buying, since that is the only strategy recommendation agents have given a history of always buying. This phenomenon occurs because there is a crash in strategy table diversity as agents lose strategy tables by imitating others.

\begin{figure*}[!htp]
\captionsetup[subfigure]{justification=centering}
    \centering
    \begin{subfigure}[]{0.49\columnwidth}
        \caption{Average volatility}
        \includegraphics[width=\textwidth]{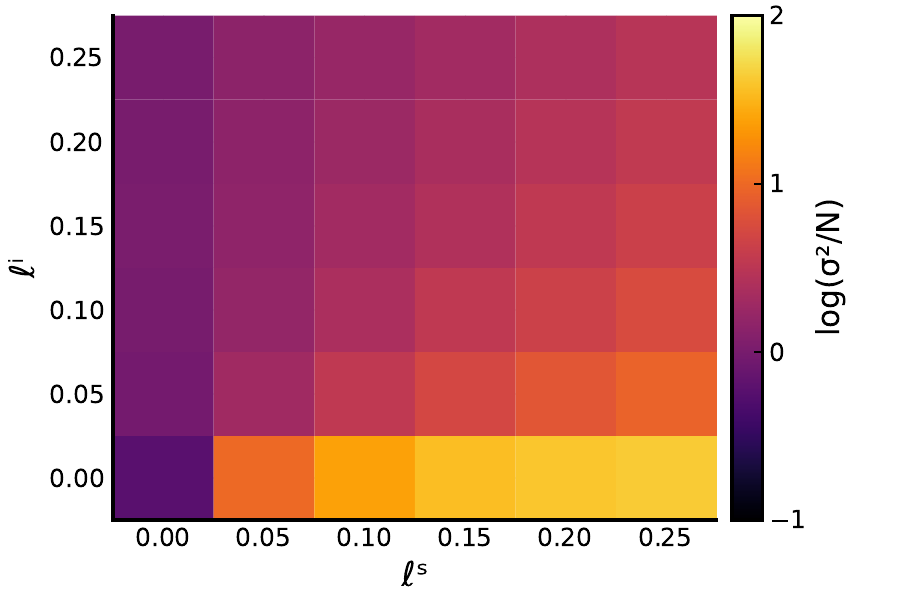} \label{fig:volatility_heatmap}
    \end{subfigure}
        \begin{subfigure}[]{0.49\columnwidth}
        \caption{Average entropy}
        \includegraphics[width=\textwidth]{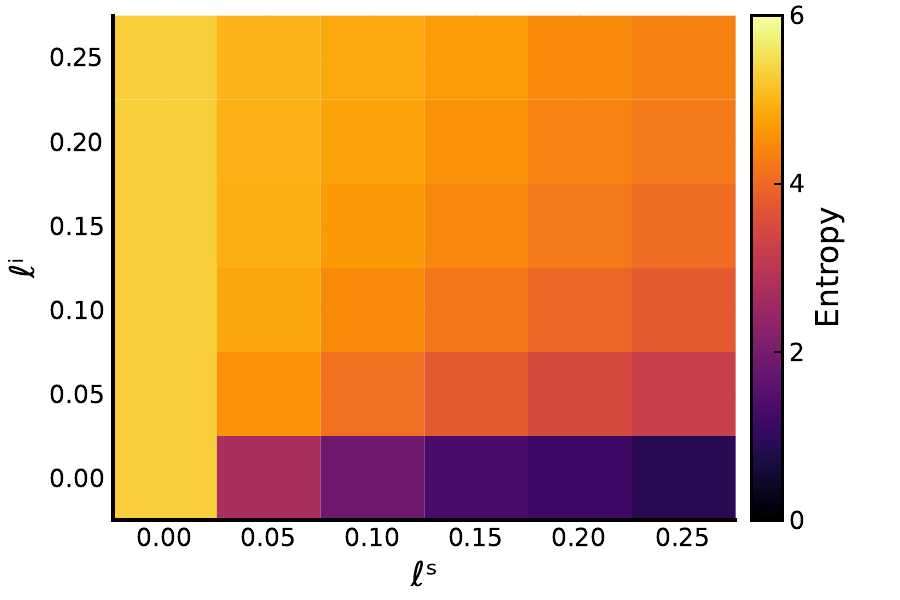} \label{fig:entropy_heatmap}
    \end{subfigure}
    \caption{Here we depict heatmaps of the log of the average volatility ($\log_{10}(\sigma^2/N$) and the average entropy when there is social and individual learning rates $\ell^s$ and $\ell^i$. Here, $M = 6$, $N = 101$, and $S=2$.}
    \label{fig:heatmaps}
\end{figure*}

To see the relationship between the diversity of strategy tables and the volatility consider Figure \ref{fig:entropy_heatmap}, which depicts the entropy (diversity) of strategy tables at the end of a game given varying degrees of social and individual learning. We can observe a relationship between high volatility, low entropy, and high social learning. The higher the level of social learning, the lower the level of entropy, since social learning undermines strategy table diversity in the population. This in turn leads to higher levels of volatility and thus less efficiency of the market. We also observe that individual learning can mitigate the negative effects of social learning, since individual learning fosters strategy table diversity.

\begin{figure*}[!htp]
\captionsetup[subfigure]{justification=centering}
    \centering
    \begin{subfigure}[]{0.49\columnwidth}
        \caption{No individual or social learning}
        \includegraphics[width=\textwidth]{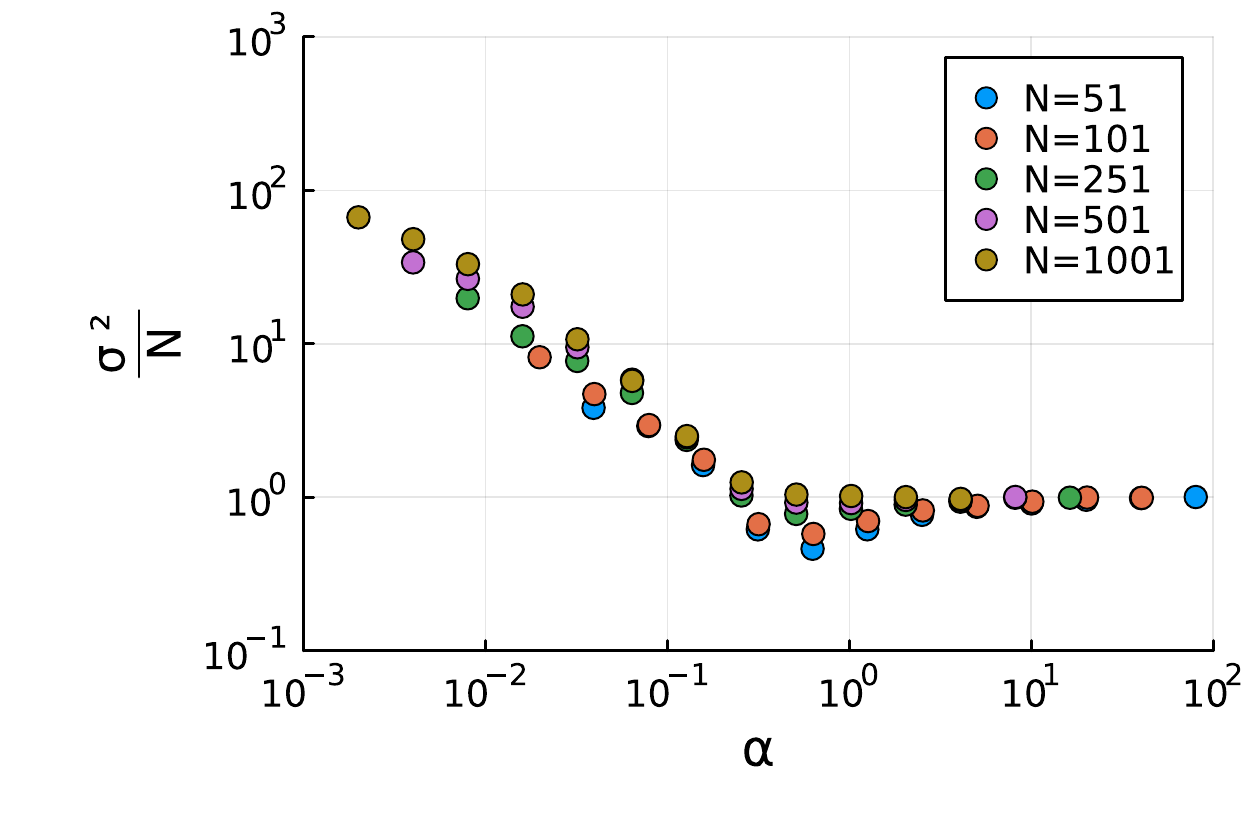} \label{fig:var}
    \end{subfigure}
    \begin{subfigure}[]{0.49\columnwidth}
        \caption{Only individual learning}
        \includegraphics[width=\textwidth]{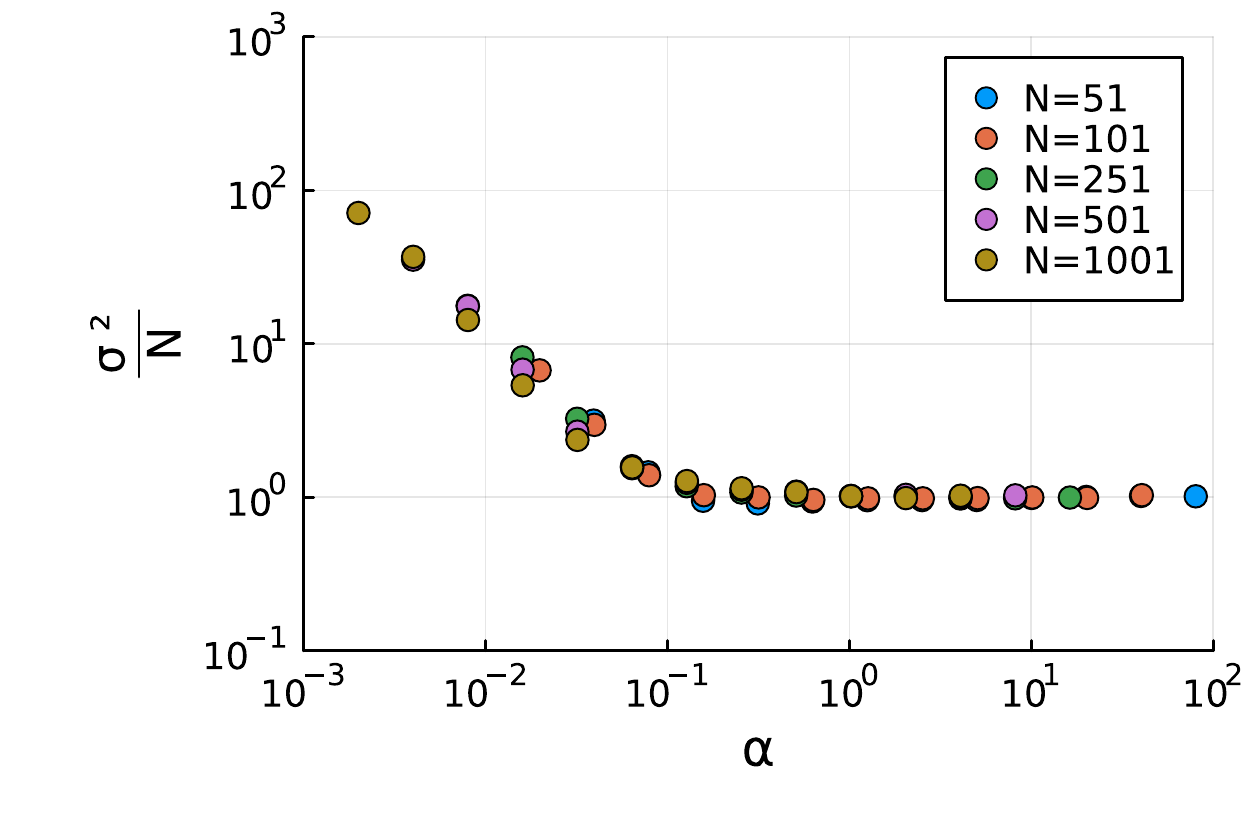} \label{fig:var_ind_learn}
    \end{subfigure}
    \begin{subfigure}[]{0.49\columnwidth}
        \caption{Only social learning}
        \includegraphics[width=\textwidth]{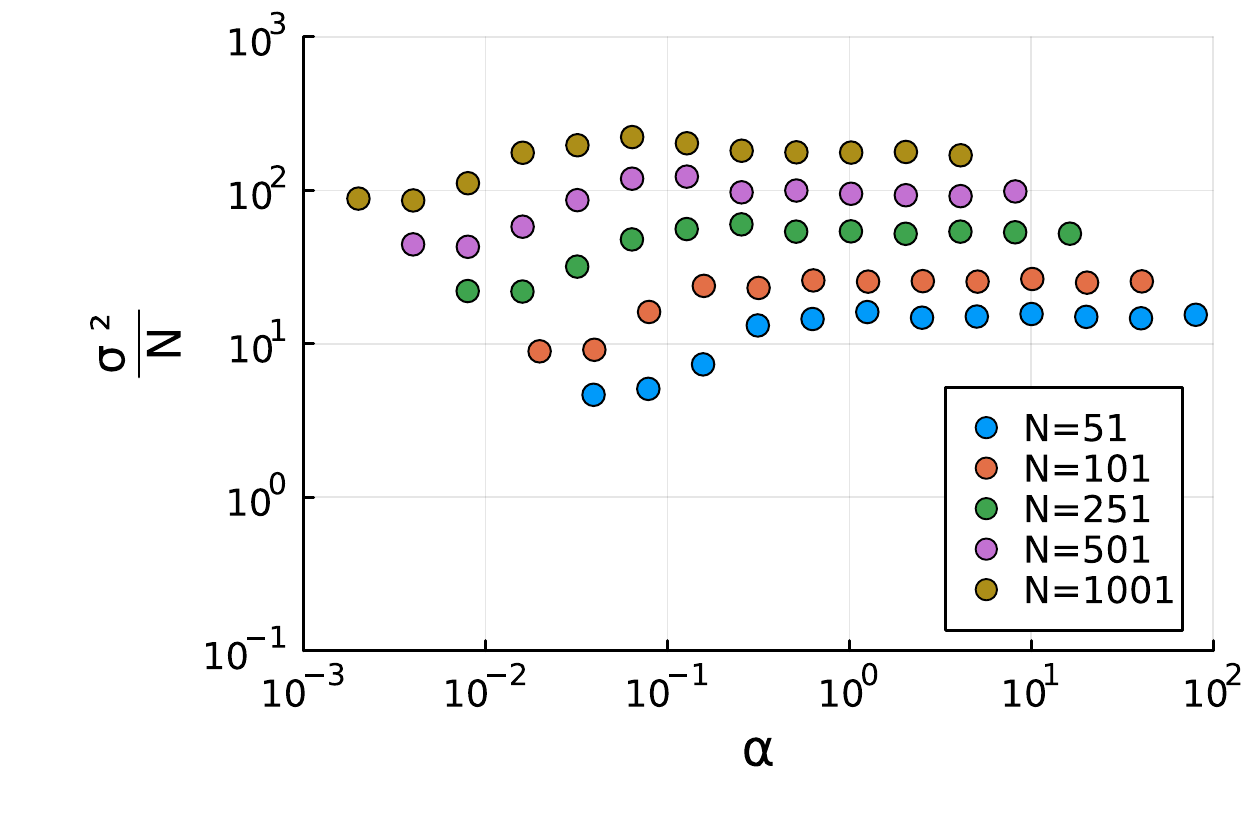} \label{fig:var_soc_learn}
    \end{subfigure}
    \begin{subfigure}[]{0.49\columnwidth}
        \caption{Individual and social learning}
        \includegraphics[width=\textwidth]{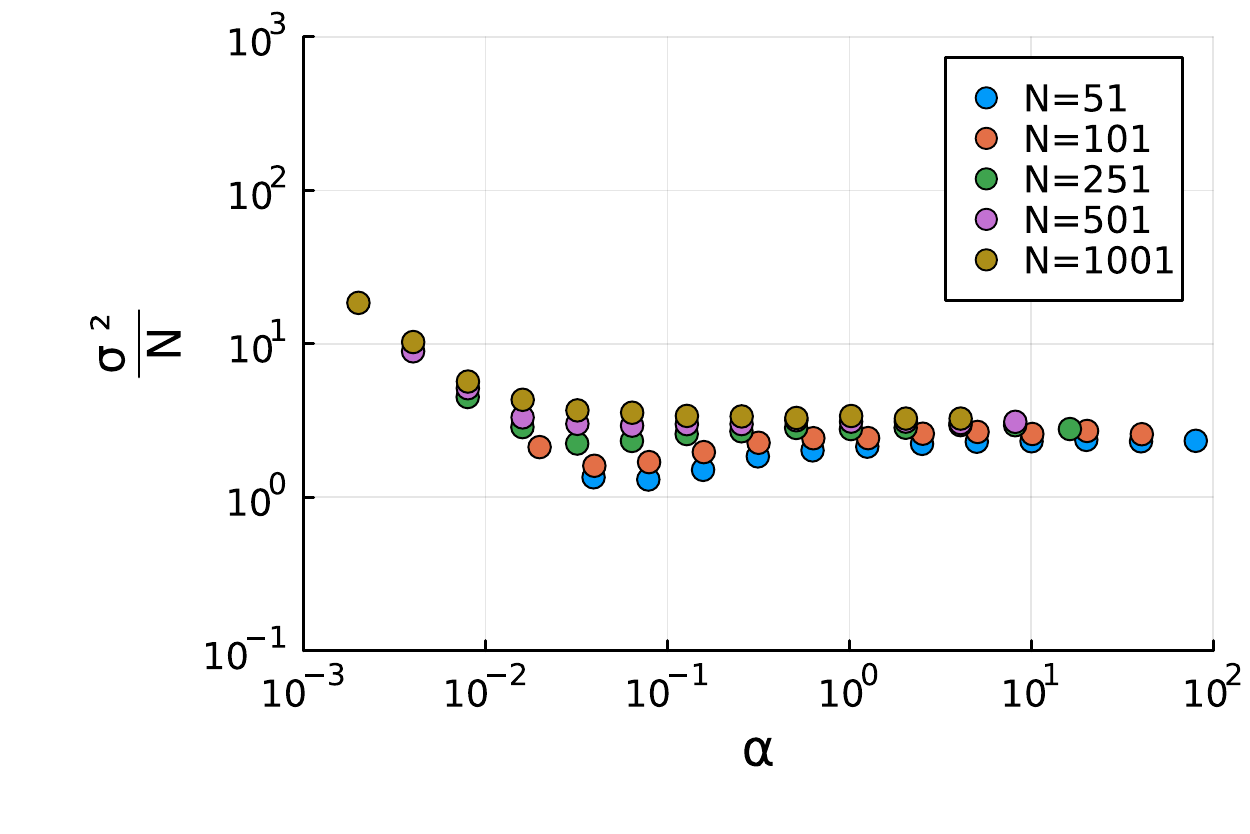} \label{fig:var_ind_soc_learn}
    \end{subfigure}
    \caption{Here we plot the normalized volatility $\sigma^2/N$ for no individual or social learning, only individual learning, only social learning, and both individual and social learning in panels a-d, respectively. Here, social learning occurs across a complete graph, $\ell^i = 0.1$ when there is individual learning, $\ell^s = 0.1$ when there is social learning, $N \in \{ 51, 101, 251, 501, 1001 \}$, $M \in \{ 1 \ldots, 12 \}$, and $S=2$. Simulations were run for $500$ turns and averaged over $20$ realizations.}
    \label{fig:volatility}
\end{figure*}

\begin{figure}
    \centering
    \includegraphics[width=0.49\columnwidth]{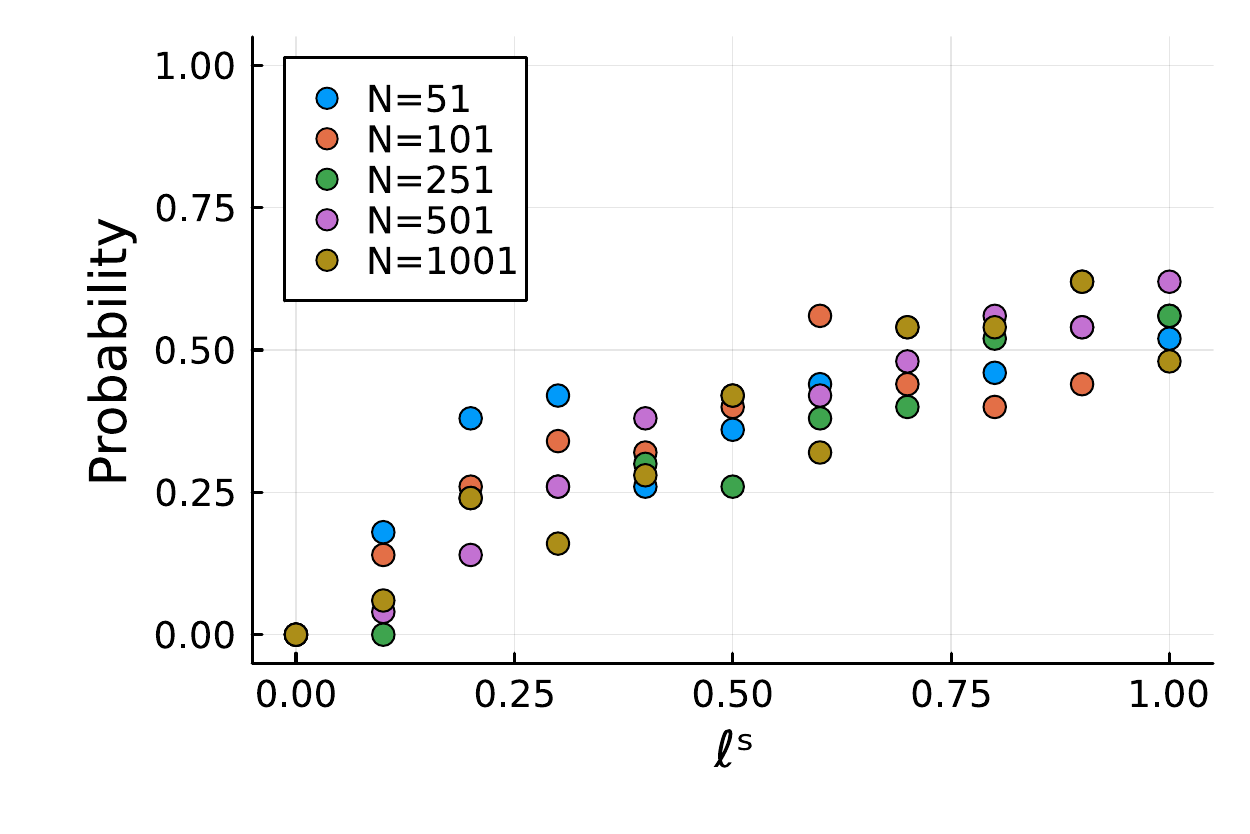}
    \caption{Probability the population becomes locked into either buying or selling under social learning. Here, $\ell^s \in \{0 \ldots, 1\}$, $N \in \{ 51, 101, 251, 501, 1001 \}$, $M=6$, and $S=2$. Simulations are run for $500$ turns and averaged over $50$ simulations.}
    \label{fig:prob_locked}
\end{figure}

Since the normalized attendance variance or volatility $\sigma^2/N$ is a key macroscopic observable of the system, we plot it vs.\ the control parameter $\alpha$ in Figure \ref{fig:volatility}. This figure compares a system with no individual or social learning, only individual learning, only social learning, and both individual and social learning. The case of no individual or social learning, depicted in Figure \ref{fig:var}, acts as a baseline for our other results. In this scenario, volatility, and thus the inefficiency of the system, decreases as we increase $\alpha$ from a low level. The system then goes through a phase transition from a symmetric phase to an asymmetric phase around $\alpha \approx 0.3374$ as has been uncovered in the literature \cite{yeung12}, after which volatility increases and levels off. Thus, volatility reaches a minimum for an intermediate $\alpha$. However, there is no such minimum for the Minority Game with individual learning (Figure \ref{fig:var_ind_learn}). Volatility simply decreases as $\alpha$ increases and levels off at approximately $1$. However, the decline is faster than when there is no such learning. Thus, for sufficiently low $\alpha$, individual learning can result in a more efficient market. When there is only social learning (Figure \ref{fig:var_soc_learn}), we observe a separation of the volatilities by population size $N$. Volatility decreases initially, but then increases before levelling off. Volatility is also much higher than when there is no individual or social learning and when there is individual learning, which aligns with the time series and heatmap results (Figures \ref{fig:timeseries} and \ref{fig:heatmaps}). Social learning results in high volatility, and this is worse the larger the population. The higher volatility from larger $N$ seems somewhat counter-intuitive, since a larger population should be more robust than a smaller one to loss of strategy diversity. However, this fact is the cause of the result. In the standard Minority Game, the time average of $A(t)$ is approximately zero due to symmetry. However, since the population can become locked into playing one strategy, volatility can go to zero in the long run for social learning as all agents make the same action every turn. This result is demonstrated in Figure \ref{fig:prob_locked}, which plots the probability that the population becomes locked into either always buying or selling. For low learning rate $\ell^s$, this probability increases as we decrease $N$. The probabilities of becoming locked into a single action increase as $\ell^s$. These probabilities for different $N$ become similar, since the outcome becomes noisier.

Though social learning can undermine efficiency in the market, allowing individuals to also learn individually can rescue it. Figure \ref{fig:var_ind_soc_learn} depicts this result for varying $\alpha$. Further, when $\alpha$ is low, volatility can be lower than the other three cases (Figures \ref{fig:var}-\ref{fig:var_soc_learn}). Social learning here is improving agents' strategy tables, and yet the negative frequency dependence and loss of diversity are being alleviated by the presence of individual learning. However, for higher $\alpha$, the cases with no individual and social learning and individual learning both have lower volatility.

\begin{figure*}[!ht]
\captionsetup[subfigure]{justification=centering}
    \centering
    \begin{subfigure}[]{0.49\columnwidth}
        \caption{No individual or social learning, $S=3$}
        \includegraphics[width=\textwidth]{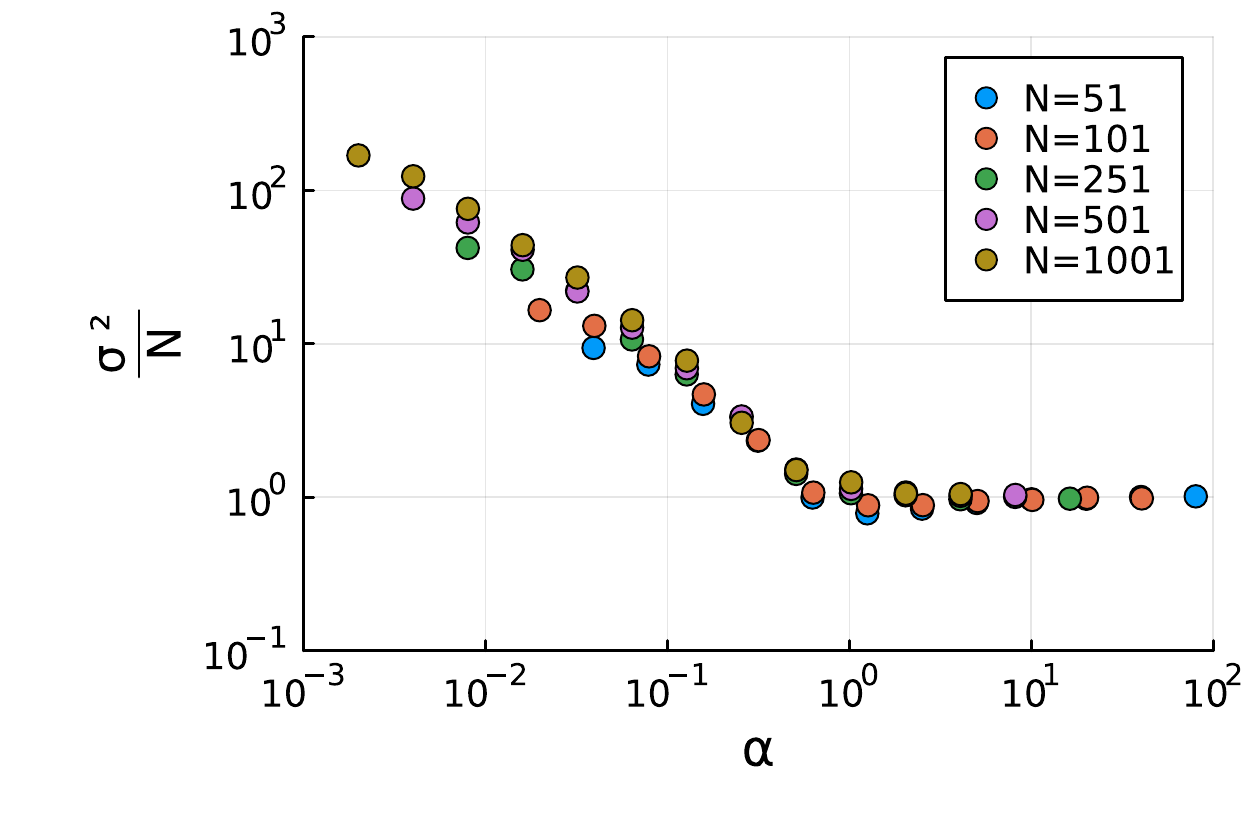}
    \end{subfigure}
        \begin{subfigure}[]{0.49\columnwidth}
        \caption{No individual or social learning, $S=4$}
        \includegraphics[width=\textwidth]{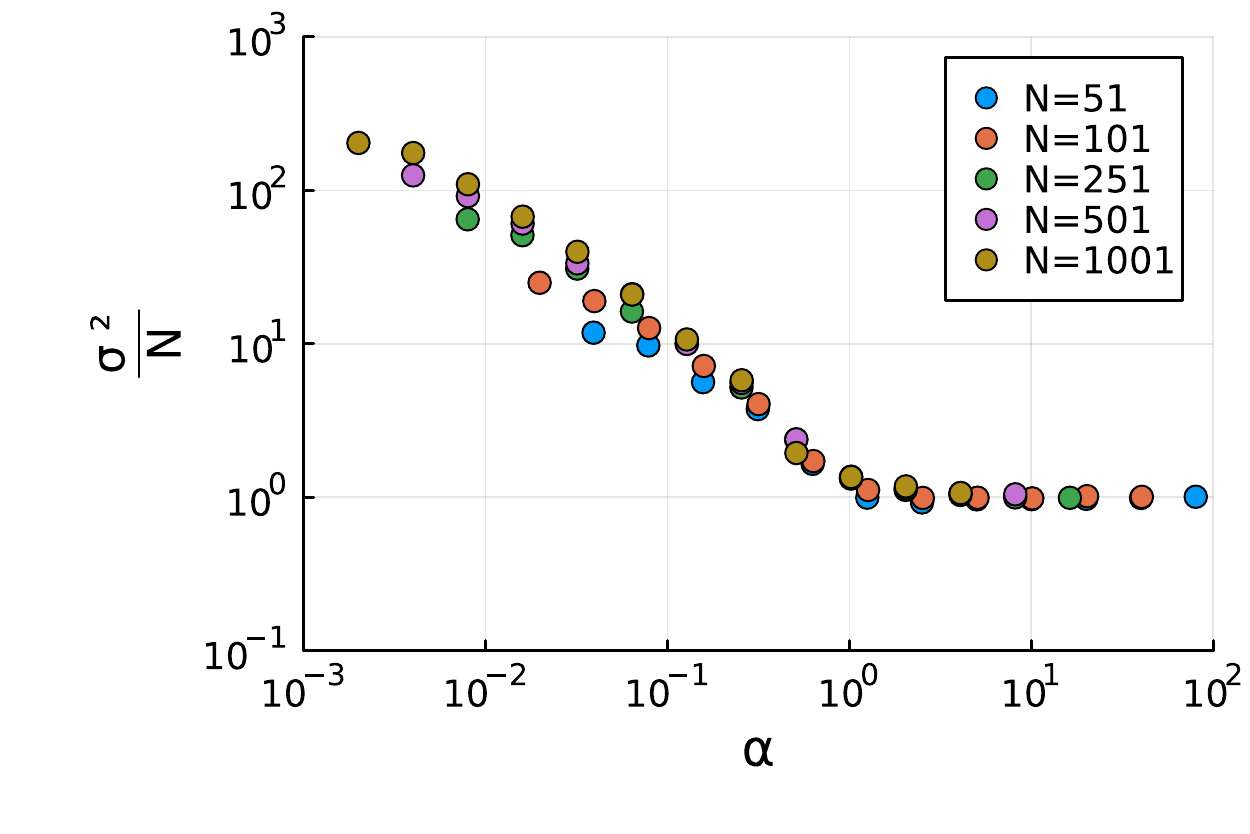}
    \end{subfigure} \\
    \begin{subfigure}[]{0.49\columnwidth}
        \caption{Only social learning, $S=3$}
        \includegraphics[width=\textwidth]{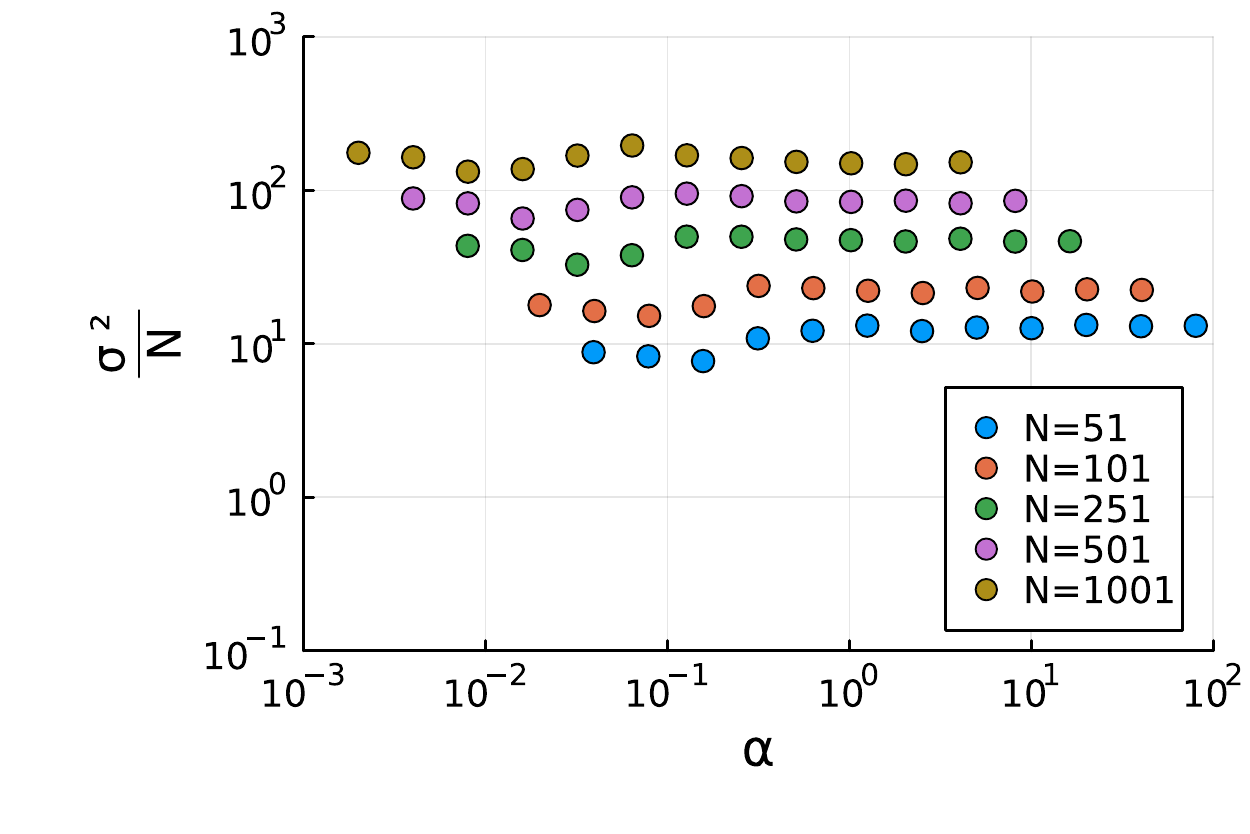}
    \end{subfigure}
        \begin{subfigure}[]{0.49\columnwidth}
        \caption{Only social learning, $S=4$}
        \includegraphics[width=\textwidth]{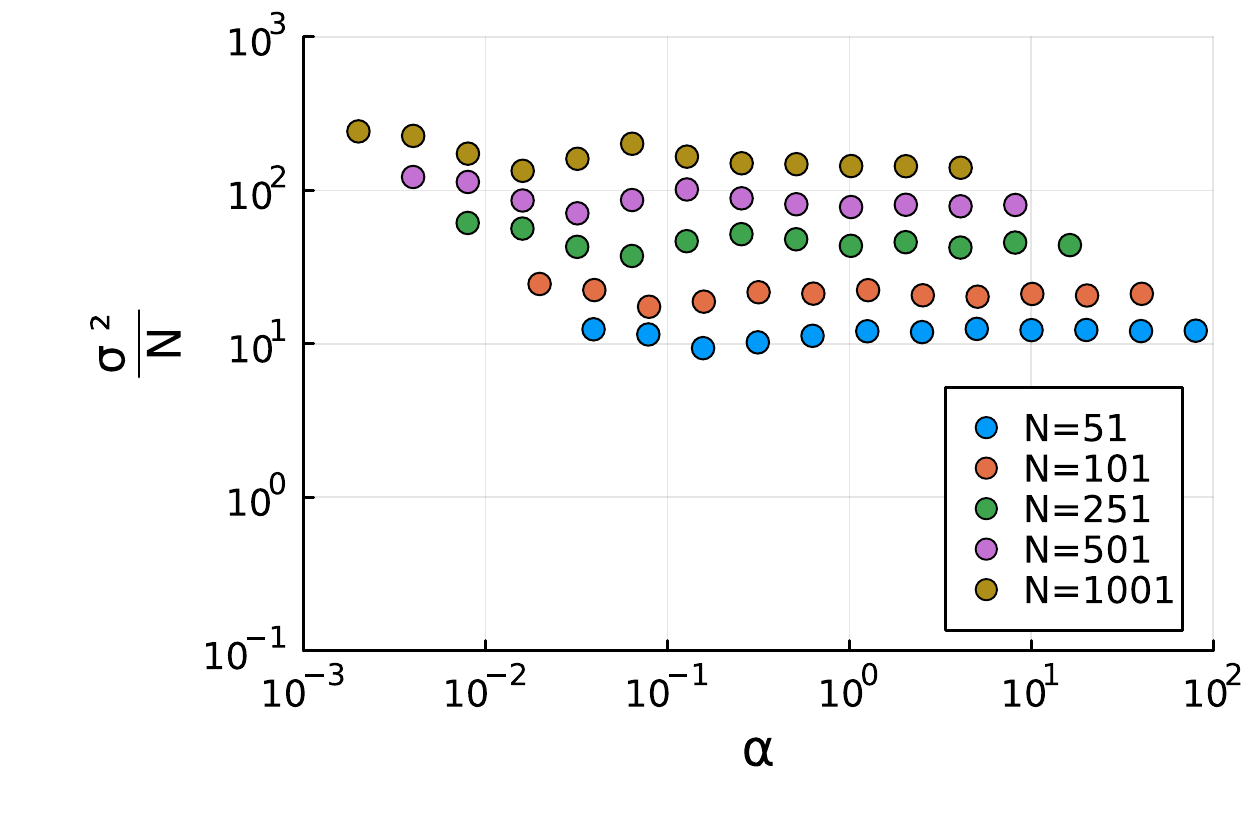}
    \end{subfigure}
    \caption{Here we plot the normalized volatility $\sigma^2/N$ for no individual or social learning and for social learning on a complete graph with $S \in \{ 3, 4 \}$. Here, $\ell^s = 0.1$, $N \in \{ 51, 101, 251, 501, 1001 \}$, and $M \in \{ 1 \ldots, 12 \}$. Simulations were run for $500$ steps and averaged over $20$ realizations.}
    \label{fig:volatility_S}
\end{figure*}

To explore the impact of further strategy tables per agent under social learning, we ran simulations for $S \in {3,4}$ as depicted in Figure \ref{fig:volatility_S}. Relative to the case where there is no individual or social learning, social learning has higher volatility. The difference, however, is less pronounced than it is when the number of strategy tables is smaller. Since there are more strategy tables per agent, diversity of strategy tables can be better maintained the greater $S$. This, however, only somewhat mitigates the harm to the market by social learning.

\begin{figure*}[!ht]
\captionsetup[subfigure]{justification=centering}
    \centering
    \begin{subfigure}[]{0.49\columnwidth}
        \caption{Only social learning, $d=0.5$}
        \includegraphics[width=\textwidth]{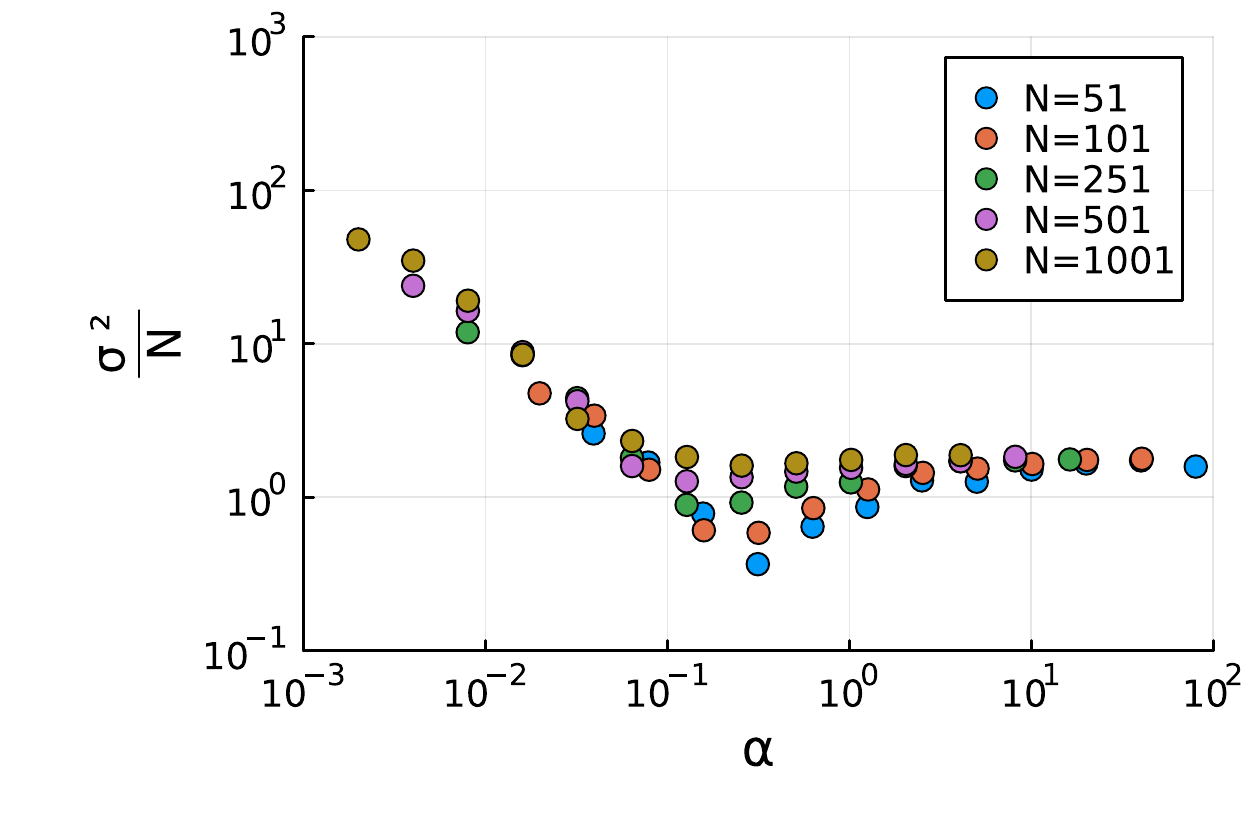}
    \end{subfigure}
    \begin{subfigure}[]{0.49\columnwidth}
        \caption{Individual and social learning, $d=0.5$}
        \includegraphics[width=\textwidth]{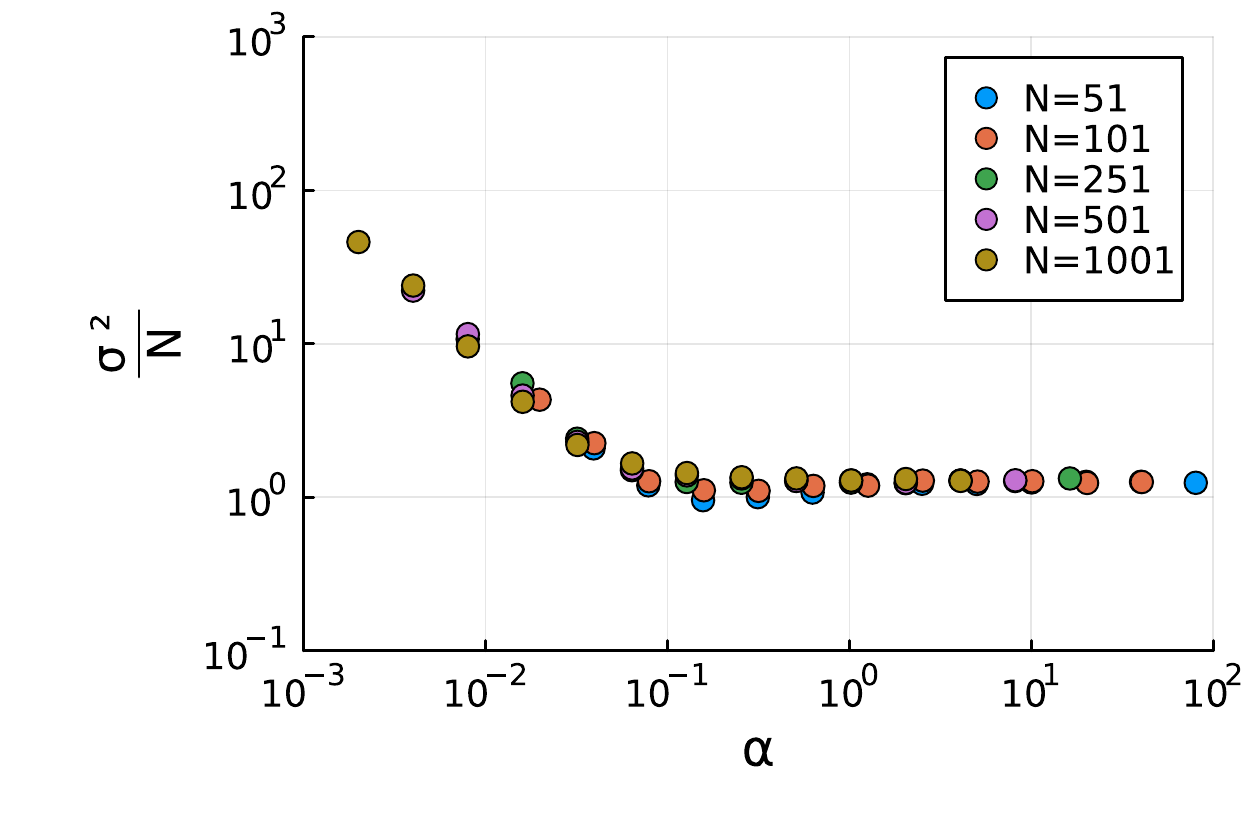}
    \end{subfigure} \\
    \begin{subfigure}[]{0.49\columnwidth}
        \caption{Only social learning, $d=1$}
        \includegraphics[width=\textwidth]{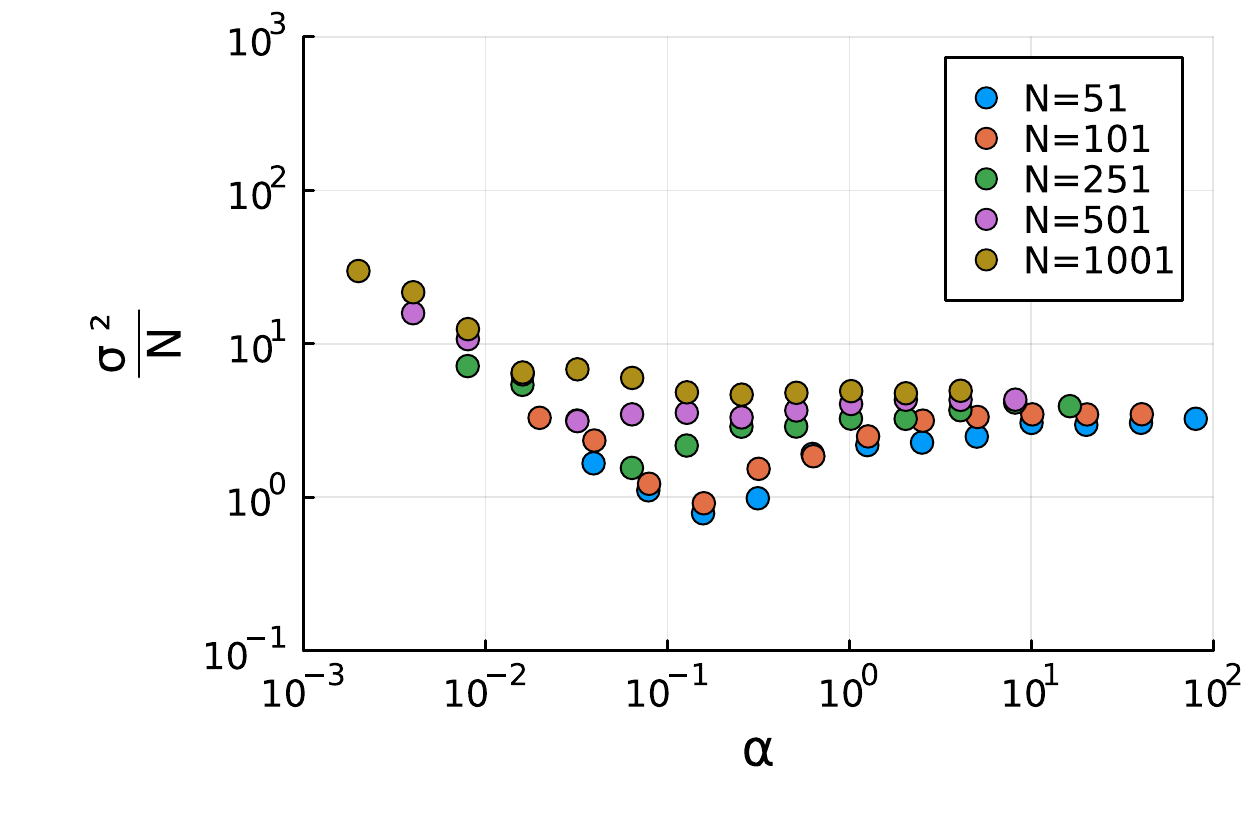}
    \end{subfigure} 
    \begin{subfigure}[]{0.49\columnwidth}
        \caption{Individual and social learning, $d=1$}
        \includegraphics[width=\textwidth]{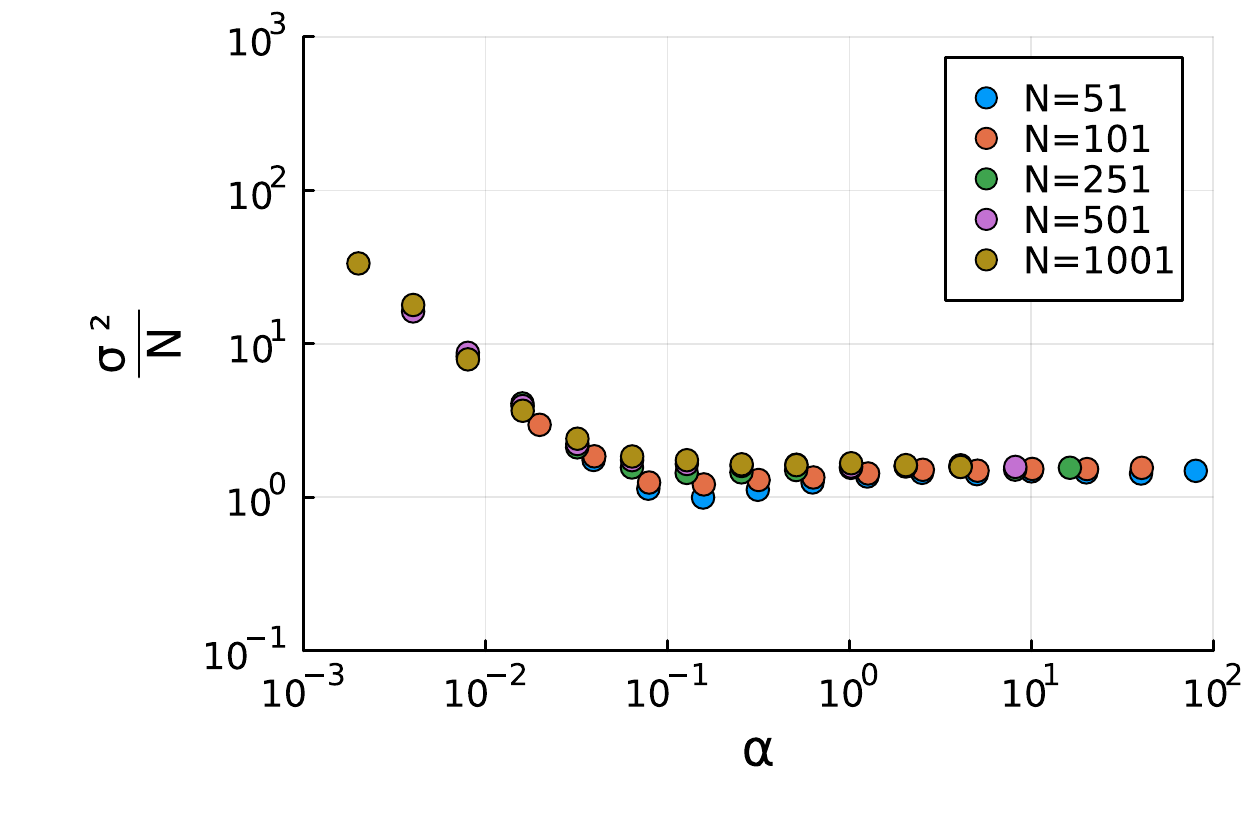}
    \end{subfigure} \\
    \begin{subfigure}[]{0.49\columnwidth}
        \caption{Only social learning, $d=2$}
        \includegraphics[width=\textwidth]{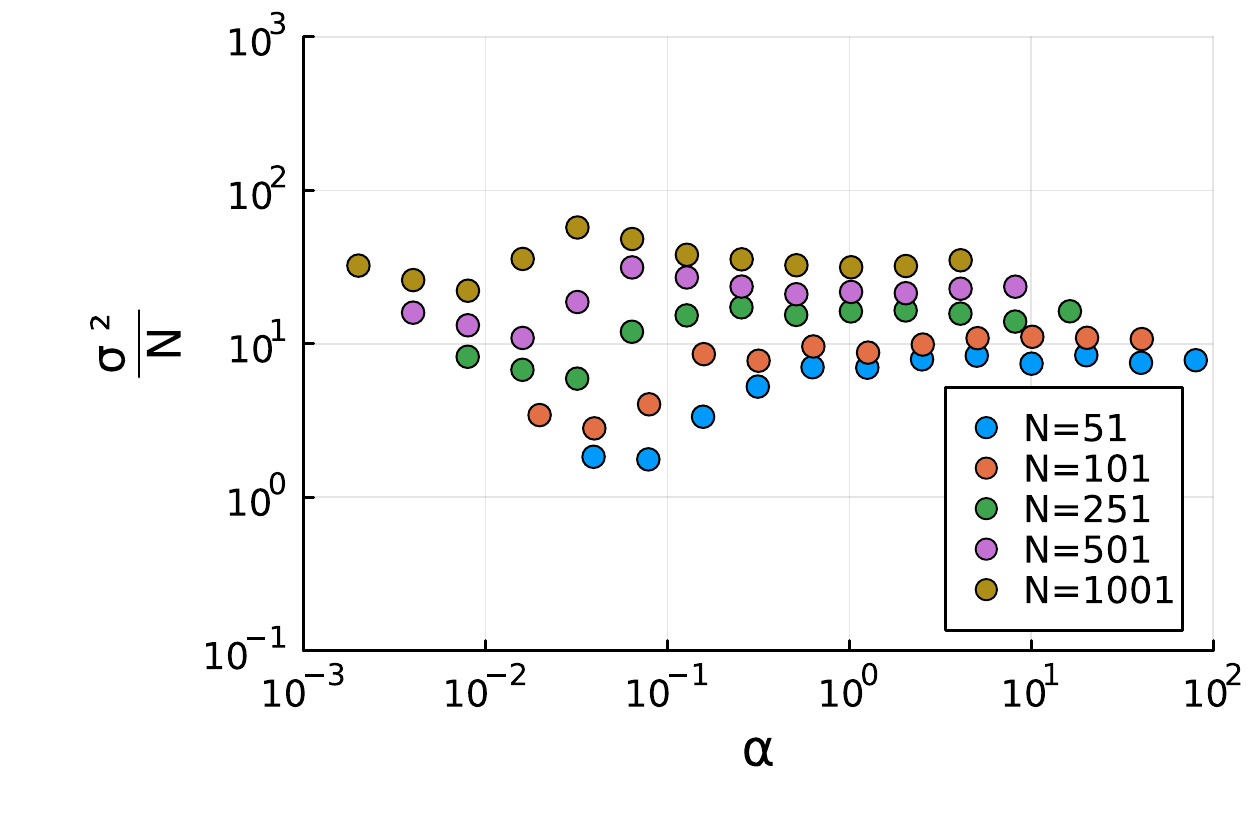}
    \end{subfigure}
    \begin{subfigure}[]{0.49\columnwidth}
        \caption{Individual and social learning, $d=2$}
        \includegraphics[width=\textwidth]{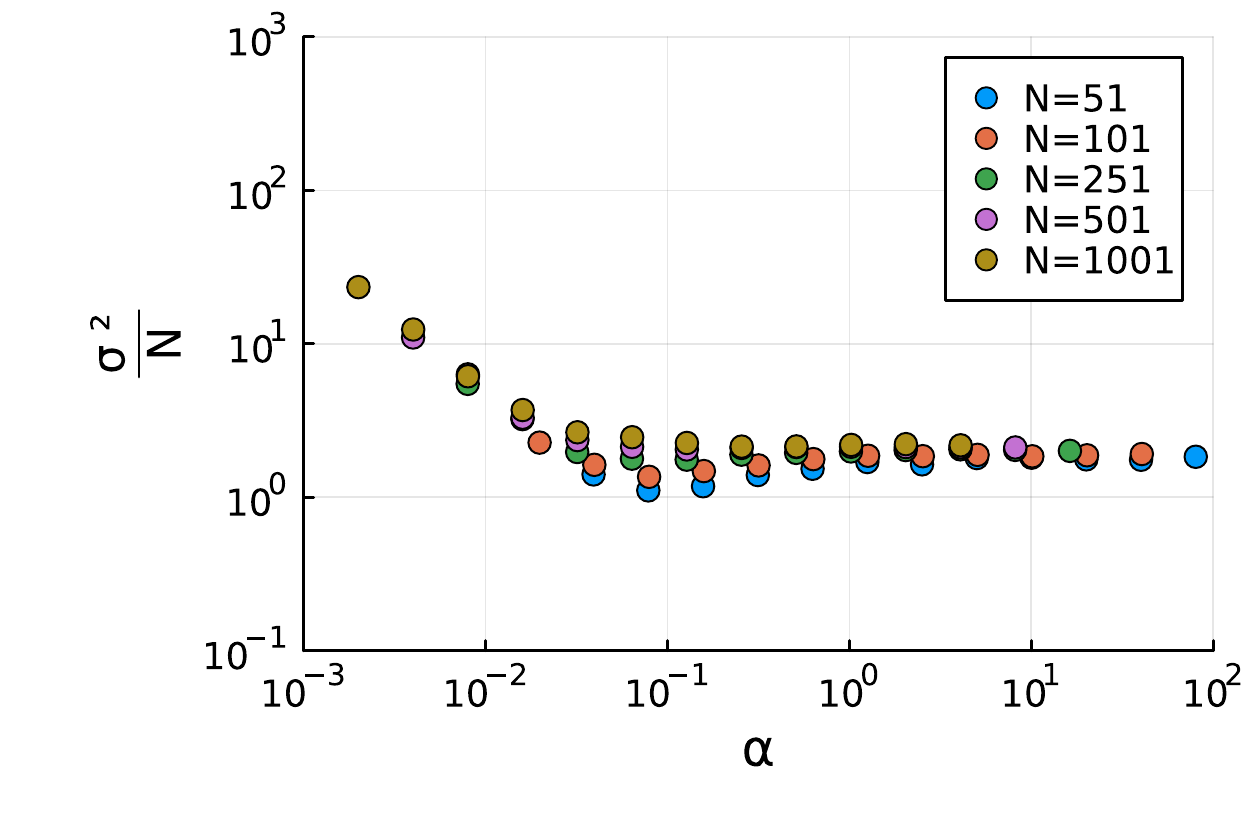}
    \end{subfigure}
    \caption{Here we plot the normalized volatility $\sigma^2/N$ for only social learning (left column) and individual and social learning (right column) on a random graph with varying mean node degrees $d \in \{ 0.5, 1, 2 \}$. Here, $\ell^i = 0.1$ (for the case of individual learning), $\ell^s = 0.1$, $N \in \{ 51, 101, 251, 501, 1001 \}$, $M \in \{ 1 \ldots, 12 \}$, and $S=2$. Simulations were run for $500$ steps and averaged over $20$ realizations.}
    \label{fig:volatility_random_graph}
\end{figure*}

Figure \ref{fig:volatility_random_graph} depicts the results for when there is only social learning and it occurs on a random graph. Even though the graph is not complete the dynamics are still quite similar qualitatively to those of a complete graph when the mean node degree is two ($d=2$). We surmise then that successful strategies can still spread through the population thereby undermining their efficacy and decreasing strategy table diversity in the population. Random graphs do suppress this effect to some degree as can be seen by comparing these figure to Figure \ref{fig:var_soc_learn}, though not sufficiently to recover the case of no social learning (Figure \ref{fig:var}). However, for $d=0.5$ and $d=1$, we observe a qualitative shift in the outcome with the results more similar to the other scenarios than the scenario with only social learning. With fewer edges by which individuals could learn, learning is sufficiently suppressed thereby increasing the efficiency of the market. When individual learning is present (right column of Figure \ref{fig:volatility_random_graph}), the impacts of individual learning dominates the outcomes, and the results are not dissimilar from those with different $d$ and those of the complete graph (Figure \ref{fig:var_ind_soc_learn}).

\section{Discussion}

Here we have explored the effects of learning on the efficiency of financial markets as modelled by the Minority Game. In a sense, the Minority Game model that includes strategy tables already features learning, since agents learn about the performance of their strategy tables. Additionally, other forms of learning have been previously studied in the Minority Game such as searching behaviour in which individuals search for information from others \cite{zhang15}. In our model, however, agents can learn new strategy tables by either copying from others or invention. We find that social learning generally undermines efficiency in markets due to the negative frequency dependence of the payoffs of strategies. Good strategies are imitated, which undermines their efficacy. This type of "self-defeating ecosystem" has been observed previously in the Minority Game and other systems \cite{batten07}. And similar results have been observed due to herding behaviour, wherein agents imitate the most informed agent \cite{cajueiro06}. Conversely, social learning has been shown to be beneficial through the ``wisdom of the crowds" phenomenon. Whether there is wisdom from the crowd or not can depend on the nature of the task. Maladaptive herding is generated by challenging tasks or when there is great uncertainty while the ``wisdom of the crowds" emerges from less challenging and more certain tasks \cite{toyokawa19}. Therefore, we would expect --- and here we have observed --- that social learning is harmful in the Minority Game as predicting the minority is a difficult task with high uncertainty.

Though we have shown that social learning can undermine efficiency in the market, it can be beneficial (at least in the short term) for individual investors. An agent who imitates a successful strategy will benefit until too many other have also adopted it. This myopic benefit of social learning in our model may provide some concordance with other research that has shown how imitation from social networks can be beneficial to investors \citep{cohen08,koochakzadeh12,wang14}. If the network is small relative to the total number of investors and thus good strategies cannot effectively spread through the wider population of investors, then social learning could be beneficial long term for this coterie of investors.

There are many possible future directions this line of research could take. For one, we have not assumed any costs associated with either social or individual learning. Since social imitation can crowd out information production \citep{han13}, exploration of such costs could be a valuable extension to this work. Additionally, we have assumed that agents know the true historic success of others' strategy tables. This assumption can be justified by noting that an agent could evaluate the historic success of any strategy it copies via its own memory. However, we could relax this assumption. On a related point, previous research has incorporating beliefs into the Minority Game \cite{zhong14}. In an extension to our model, agents could incorrectly estimate the success of others' strategies through biased beliefs. Another future direction is biased imitation. For example, though our agents each make only one trade/action a turn, imitation of investing strategies can be biased by volume of trade \cite{simon12}, via a self-enhancing transmission bias \citep{han16}, or through homophilic imitation of agents with cultural or social traits like their own \citep{morsky17}. Biased transmission is the primary force in cultural evolution \cite{henrich01a}. Though our network approach could be considered an example of this, explicitly modelling the prestige of information and agents, and how that affects imitation and the spread of strategies. Prestige processes are important in cultural transmission and can improve the quality of information spread \cite{henrich01b}. However, in the Minority Game, this likely would undermine the efficiency of the market. A thorough study of prestige, financial gurus, and other factors biasing the spread of cultural traits applied to the Minority Game would be an interesting topic of future research.

\subsection*{Acknowledgments}

This work was supported in part by the First Year Assistant Professor Award from the Council on Research and Creativity of Florida State University.

\subsection*{Code and data availability}
Code to run numerical simulations is available at https://github.com/bmorsky/minority-games.

\bibliography{minority}
\bibliographystyle{plain}

\end{document}